\documentstyle[12pt,aaspp4]{article}

\begin{document}

\title{MCG+00-32-16:  An Irregular Galaxy Close to the Lowest Redshift Absorber on the 3C$\;$273 Line of Sight}

\author{G. Lyle Hoffman}
\affil{Dept. of Physics, Lafayette College, Easton, PA  18042; hoffmang@nova.phys.lafayette.edu}

\author{Nanyao Y. Lu}
\affil{IPAC, Mail Stop 100-22, California Institute of Technology, Pasadena, CA 91125; lu@ipac.caltech.edu}

\author{E.E. Salpeter}
\affil{Center for Radiophysics and Space Research, Cornell University, Ithaca, NY  14853; salpeter@spacenet.tn.cornell.edu}

\author{Bryan M. Connell}
\affil{Dept. of Physics, Lafayette College, Easton, PA  18042; connellb@lafayette.edu}

\and

\author{Ren\'{e} Fromhold-Treu}
\affil{Dept. of Physics, Lafayette College, Easton, PA  18042}

\received{10/1/97}
\accepted{1/13/98}
\slugcomment{to appear in {\em Astrophysical Journal}}

\begin{abstract}

MCG+00-32-16 is the galaxy closest in position-velocity space to the lowest redshift Ly$\alpha$ absorber along the line-of-sight to the quasar 3C$\;$273.
Its projected separation is $204 ( d / 19 {\rm Mpc} )$ kpc, where d is the distance from the Milky Way to the galaxy, and the redshift difference is only 94 km ${\rm s}^{-1}$; \ion{H}{1}$\;$1225+01 is slightly closer in projected separation to the absorber, but has a greater redshift difference.
We present \ion{H}{1} synthesis array mapping and CCD photometry in B and R for MCG+00-32-16.
The \ion{H}{1} disk is rotating in such a way that the side of the galaxy closer to the sight-line to the quasar has the larger velocity difference from the absorber. 
The absorber may be a ``failed dwarf'' member of a poor galaxy group of which MCG+00-32-16 and \ion{H}{1}$\;$1225+01 are the only members to have formed stars.
\end{abstract}

\keywords{Galaxies: Individual: MCG+00-32-16; Galaxies: Irregular; Galaxies: Intergalactic Medium; Galaxies: Photometry; Galaxies: Quasars: Absorption Lines; Radio Lines: Galaxies}

\section{Introduction}

Even before observational data became available, there were predictions of intergalactic Ly$\alpha$ absorption, either continuous (Gunn \& Peterson \markcite{GP65}1965) or in discrete lines from galaxy halos or clusters of galaxies (Spitzer \markcite{Sp56}1956; Bahcall \& Salpeter \markcite{BS65}1965; Bahcall \& Spitzer \markcite{BS69}1969).
The observed damped wing Ly$\alpha$ absorption lines, rare in occurrence but high in \ion{H}{1} column density (of order $10^{21}$ atoms ${\rm cm}^{-2}$), are generally thought to be produced by inner protogalaxy disks, but there is on-going controversy about the very abundant Ly$\alpha$ forest lines with very low \ion{H}{1} column density ($\lesssim 10^{13}$ atoms ${\rm cm}^{-2}$).
The Ly$\alpha$ forest at high redshifts shows some velocity correlation which suggests some association with galaxies (Fernandez-Soto et al. \markcite{FLBCWY96}1996), but less so than at low redshifts (Ulmer \markcite{Ul96}1996), and the high-z forest is usually thought of in terms of density enhancements which have not yet formed equilibrium structures.
We shall not discuss models of this kind (Zhang et al. \markcite{ZAN95}1995; Hernquist et al. \markcite{HKWM96}1996; Miralda-Escud\'{e} et al. \markcite{MCOR96}1996;  Bi \& Davidsen \markcite{BD97}1997) here, but will discuss only the Ly$\alpha$ forest associated with galaxies at low (and intermediate) redshifts, excluding the smaller number of absorbers in voids (Stocke et al. \markcite{SSPDC95}1995; Shull et al. \markcite{SSP96}1996).

At least two types of locations for Ly$\alpha$ forest gas have been proposed:  (1) direct association with the outskirts of an individual normal galaxy, and (2) somewhere in a galaxy group outside of individual galaxies.
For locations of type (1), whether the association is in terms of the earlier spherical halo models or extended disks of ordinary galaxies (Maloney \markcite{Ma92}1992) or almost invisible ``failed dwarf'' or ``Cheshire Cat'' galaxies (Salpeter \markcite{Sa93}1993; Salpeter \& Hoffman \markcite{SH95}1995; Jimenez et al. \markcite{JHHP97}1997) or a collection of clouds in a halo, one needs radii of a few hundred kpc for the extension of the galaxy, 10 to 100 times larger than optical galaxy radii.
The hypothesis (2) of Ly$\alpha$ forest absorbers residing in galaxy groups (Morris \& van den Bergh \markcite{MvdB94}1994; Mo \& Morris \markcite{MM94}1994) is motivated in part by the fact that the number density and typical size ($\sim 1$ Mpc) of groups is compatible with the observed number of forest lines, if the area covering factor within each group is appreciable.
For type (2) locations there is the further controversy over whether or not the absorbing gas was ejected from individual galaxies (via tidal interactions or starburst-powered outflows; Wang \markcite{Wa95}1995) or resides in a separate optically invisible dwarf galaxy (faded or failed) with its own dark matter halo.

The discovery by Hubble Space Telescope (HST) of Ly$\alpha$ absorbing clouds at low redshift along the lines of sight to a number of bright quasars (Bahcall et al. \markcite{BJSHBJ91}1991, \markcite{BJSHG92}1992a, \markcite{BJSHJ92}1992b, \markcite{BJSH93}1993; Morris et al. \markcite{MWSG91}1991; Weymann et al. \markcite{WRWMH95}1995) has led to a number of follow-up studies.
Some observations document locations (1), galaxy extensions up to $\sim 100$ kpc (Lanzetta et al. \markcite{LWB96}1996; Barcons et al. \markcite{BLW95}1995); some document locations (2), galaxy groups (Salzer \markcite{Sa92}1992; Morris et al. \markcite{MWD+93}1993; Salpeter \& Hoffman \markcite{SH95}1995; Lanzetta et al. \markcite{LWB96}1996; Bowen et al. \markcite{BBP96}1996, \markcite{BOBT97}1997a, \markcite{BPB97}1997b; Tripp et al. \markcite{TLS97}1997).
Additional studies on explicit galaxy-absorber association, which seek correlations among the absorbers' equivalent widths, impact parameters, velocity differences and the galaxies' luminosities, suggest that both (1) and (2) occur (Chen et al. \markcite{CLWB98}1998; Le Brun et al. \markcite{LBB96}1996; Pettijean et al. \markcite{PMK95}1995; Rauch et al. \markcite{RWM96}1996; Shull et al. \markcite{SSP96}1996; van Gorkom et al. \markcite{vG+93}1993, \markcite{vG+96}1996).

In those cases where a galaxy has been observed moderately close in position and velocity to a Ly$\alpha$ forest line, the sense of the galaxy's rotation curve can give additional information toward a choice between types (1) and (2), if the galaxy's major axis points roughly toward (or away from) the line-of-sight to the quasar.
\ion{H}{1} mapping by van Gorkom et al. (\markcite{vG+96}1996) uncovered two \ion{H}{1}-rich galaxies with impact parameters $< 300\;$kpc to quasar sight-lines and very small velocity differences from the absorbers, but both galaxies' major axes are almost perpendicular to the directions to the absorbers.
Barcons et al. \markcite{BLW95}(1995) have found two cases where the side of the galaxy closer to the quasar sight line also has a velocity closer to that of the forest absorption line, possibly suggesting a direct galaxy extension, but the velocity differences are still larger than expected from simple rotation.
This paper adds another rotation curve to the sample.

We discovered that MCG+00-32-16, at projected separation $204 ( d / 19 {\rm Mpc} )$ kpc and velocity difference 94 km ${\rm s}^{-1}$, was the galaxy closest in position-redshift space to the lowest redshift Ly$\alpha$ absorber on the line-of-sight to 3C$\;$273 in the course of our study of \ion{H}{1}-bearing galaxies in that neighborhood (Salpeter \& Hoffman \markcite{SH95}1995).
At that time, the only \ion{H}{1} observation of MCG+00-32-16 was the single-beam Nan\c{c}ay spectrum of Garcia et al. \markcite{GBGGP93}(1993).
We were interested in the velocity field of the galaxy, and in particular whether or not an extrapolation of the \ion{H}{1} disk observable in emission could be responsible for the absorption.
Any indications of \ion{H}{1} clouds outside of the galaxy proper would also have been of interest (van Gorkom et al. \markcite{vG+93}1993, \markcite{vG+96}1996).
In Sect. \ref{observations}, we present synthesis array mapping and CCD surface photometry in B and R for MCG+00-32-16 and undertake a rotation curve decomposition. 
We discuss various models in Sect. \ref{discuss} and end with a summary and conclusions in Sect. \ref{conclusions}.

\section{\label{observations}Observations}

\subsection{Neutral hydrogen mapping}

The \ion{H}{1} mapping of MCG+00-32-16 was conducted at the Very Large Array\footnote{The Very Large Array of the National Radio Astronomy Observatory is a facility of the National Science Foundation, operated under cooperative agreement by Associated Universities, Inc.} 
as detailed in Table \ref{obsdet}, which lists the date of the observing run, the pointing center (epoch 1950), the heliocentric velocity to which the receivers were tuned, the array configuration, the number of spectral channels and channel separation in velocity units, and the time spent on source during the observations, not including time spent on calibrators.
We have also listed the resulting beam size in arcsec, the rms noise per channel, and the $3\sigma$ detection limit for \ion{H}{1} column density in each channel.
We obtained both R and L circularly polarized data.
On-line Hanning smoothing was employed, and the observations were calibrated using sources from the VLA calibrator list.
Calibration and data-editing were accomplished using standard AIPS tasks.
Continuum was subtracted in the uv domain using the task UVBAS, and maps were made and CLEANed using IMAGR with robustness parameter 0 for the best compromise between spatial resolution and maximum signal-to-noise.
After imaging, the data cube was corrected for the VLA primary beam.

\begin{figure}
\plotone{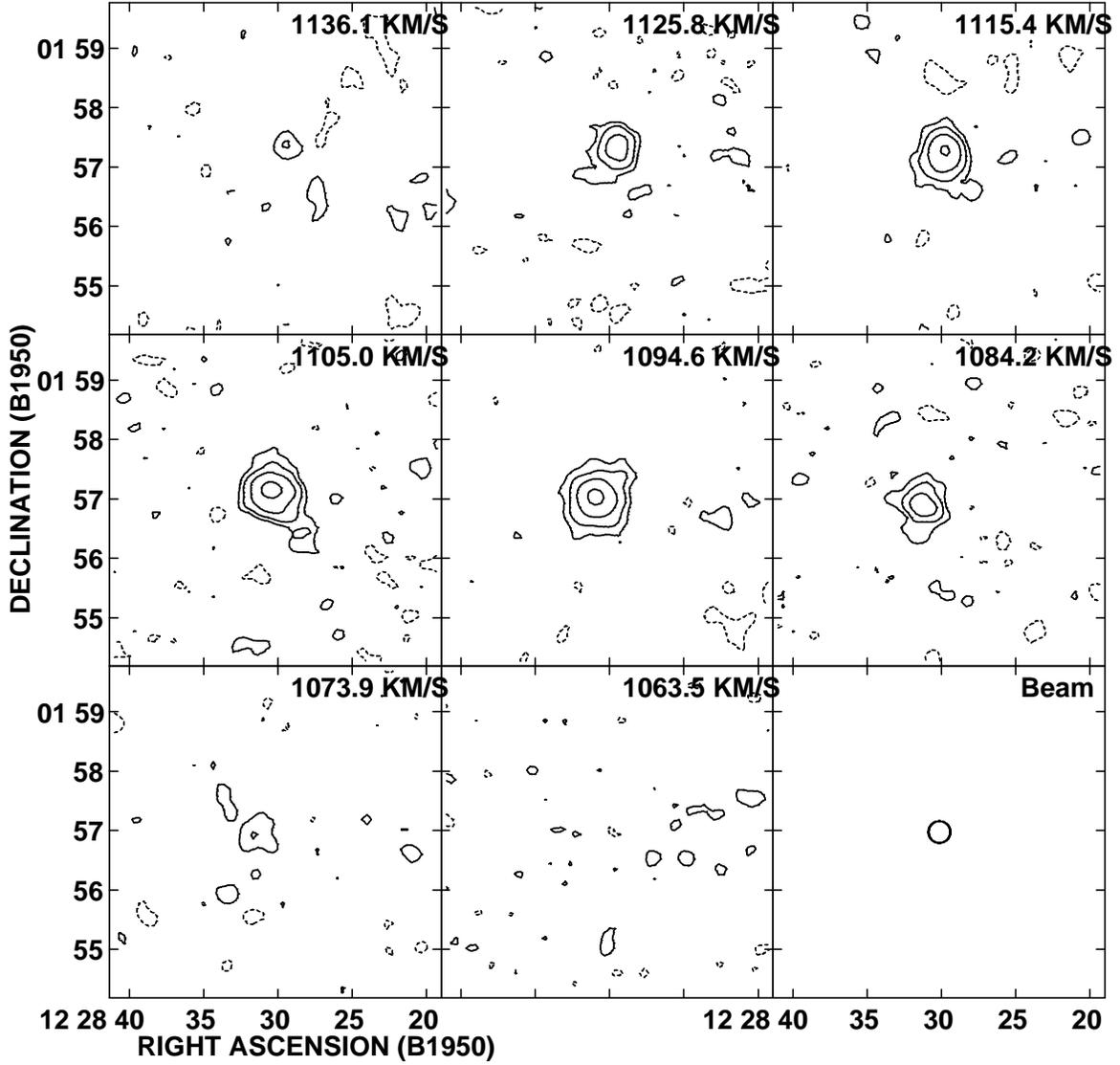}
\caption{
\label{mcgpanel}
Mosaic of channel contour maps for MCG+00-32-16, labeled by the heliocentric velocity of each channel in the upper right corner of each pane.
The beam is shown in a separate pane.
Contours are drawn at $-4.3$, 4.3, 8.6, 17, and 35$\times 10^{19}$ atoms ${\rm cm}^{-2}$.}
\end{figure}

Figure \ref{mcgpanel} shows a mosaic of the line-bearing channels of the MCG+00-32-16 data cube.
No other hints of signal were seen outside the field of view of these panes, nor in other velocity channels, even after the data were smoothed to $45 \times 45\arcsec$ resolution.
We have no evidence of extended diffuse emission nor of outlying diffuse clouds of \ion{H}{1} to our ($3 \sigma$) sensitivity limit, $\sim 1.1 \times 10^{20}\;$atoms$\;\rm{cm}^{-2}$ per velocity channel, or $\sim 3 \times 10^{19}\;$atoms$\;\rm{cm}^{-2}$ per velocity channel after smoothing.

\begin{figure}
\plotone{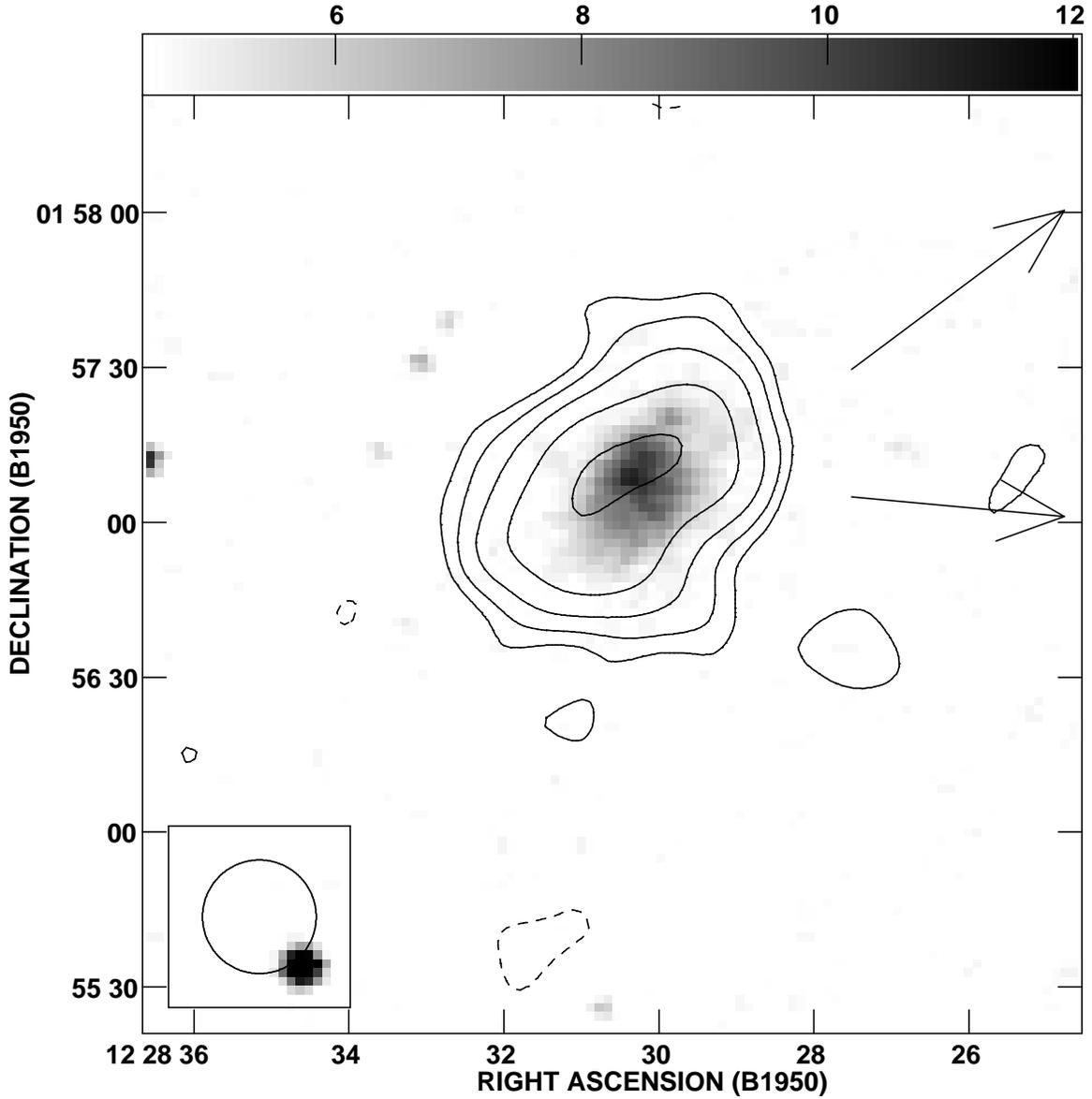}
\caption{
\label{mcgtothi}
Contours of total hydrogen emission from MCG+00-32-16, integrated over all line-bearing channels, superimposed on a grey scale image from the Digitized Sky Survey.
Contour levels are $-2.3$, 2.3, 3.4, 5.1, 7.7 and 11.6$\times 10^{20}$ atoms ${\rm cm}^{-2}$.
The upper arrow points toward 3C$\;$273, about $37\arcmin$ NW; the lower one points toward HI 1225+01, about $50\arcmin$ W.}
\end{figure}

A contour map of the total \ion{H}{1} emission (0th moment map with a flux cutoff of $- 2 \sigma$) is superimposed on a grey scale image from the Digitized Sky Survey of the optical emission from MCG+00-32-16 in Fig. \ref{mcgtothi}.
\ion{H}{1} extends, to our sensitivity, to the edge of the optical emission or a bit beyond; MCG+00-32-16 does not appear to be a dwarf with a dramatically extended \ion{H}{1} disk like DDO 154 (Hoffman et al. \markcite{HLSFLR93}1993; Carignan et al. \markcite{CBF90}1990; Krumm \& Burstein \markcite{KB84}1984) or \ion{H}{1}$\;$1225+01 (Giovanelli et al. \markcite{GWH91}1991).
However, we have some marginal evidence that our C array observations have resolved out some diffuse flux:  The total \ion{H}{1} flux integrated over the map is 1.76 Jy km ${\rm s}^{-1}$, whereas Garcia et al. \markcite{GBGGP93}(1993) found 2.3 Jy km ${\rm s}^{-1}$ using the Nan\c{c}ay single beam.
Our flux amounts to $1.5 \times 10^8 (d / 19\;{\rm Mpc} )^2\:{\rm M}_{\sun}$ where $d$ is the distance to the galaxy, assumed to be a member of the Southern Extension of the Virgo Cluster.
MCG+00-32-16 is, in fact, within the ``triple-valued region'' of the Virgo Cluster, with possible distances of 8.6, 19 or 26 Mpc in a spherical infall model which has the global value of $H_{o} = 74\;{\rm km}\;{\rm s}^{-1}\;{\rm Mpc}^{-1}$, $\Omega_{o} = 1$, and a Local Group infall velocity of $273\;{\rm km}\;{\rm s}^{-1}$.
Tully-Fisher (TF) estimates of the distance to the galaxy are inconclusive due to its small profile width and the wide scatter in the relation for irregular galaxies; however, application of the blue TF relation from Hoffman et al. (\markcite{HHS88}1988) favors the middle distance.

\begin{figure}
\plotone{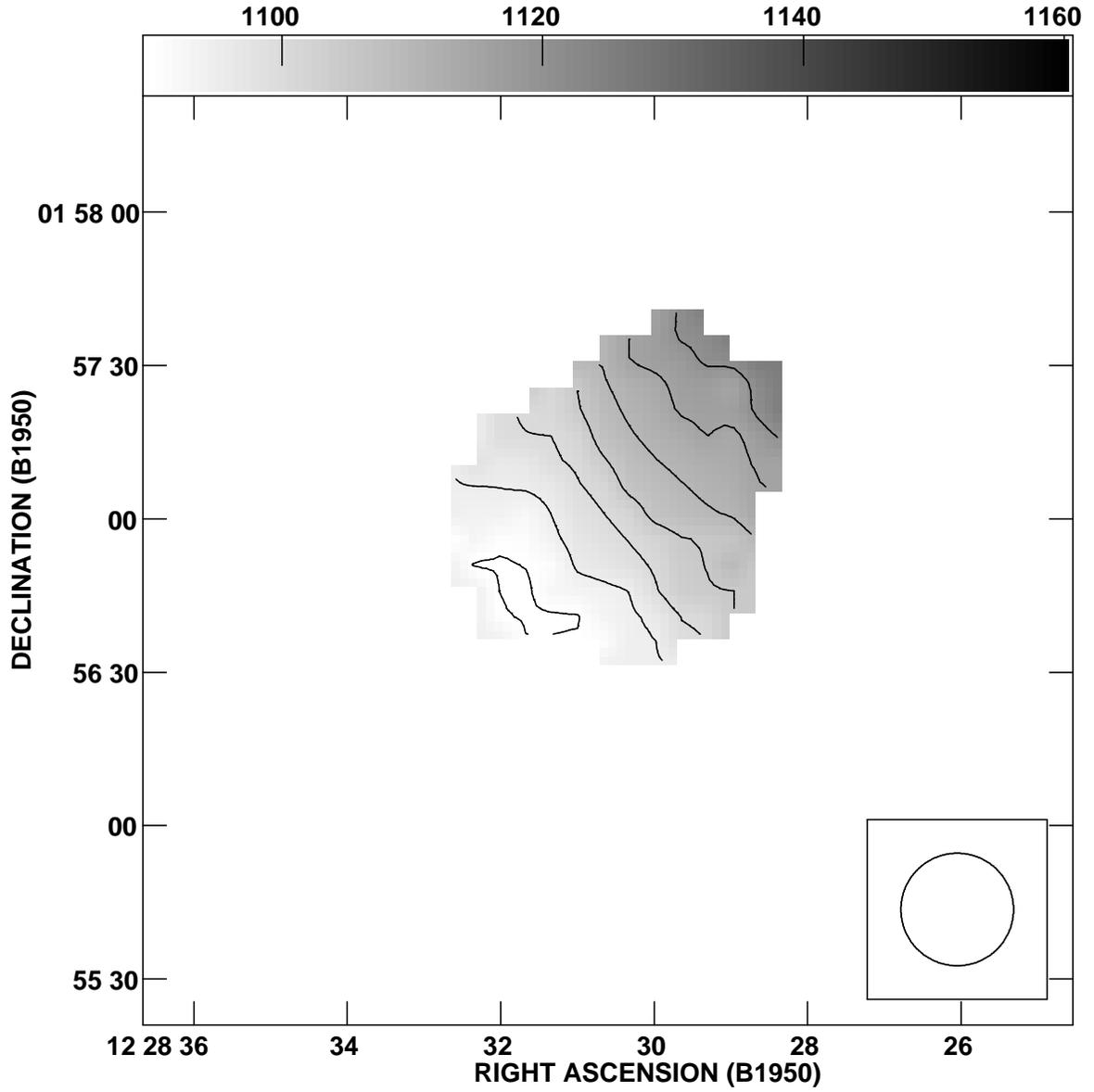}
\caption{
\label{mcgvel}
Isovelocity contours of MCG+00-32-16 superimposed on a greyscale image of the same.
The contours are drawn from 1090 to 1120 in steps of 5 km $\rm s ^{-1}$.}
\end{figure}

The velocity field of MCG+00-32-16 is shown in Fig. \ref{mcgvel}.
The galaxy appears to be in quite regular solid body rotation out to the edge of the stellar distribution, typical of many dwarf irregulars.
The line-of-sight velocity dispersion is essentially constant over the face of the galaxy; the average value is 10.7 km ${\rm s}^{-1}$, but since this is comparable to the channel spacing it should be regarded more as an upper limit than a measurement.

\begin{figure}
\plottwo{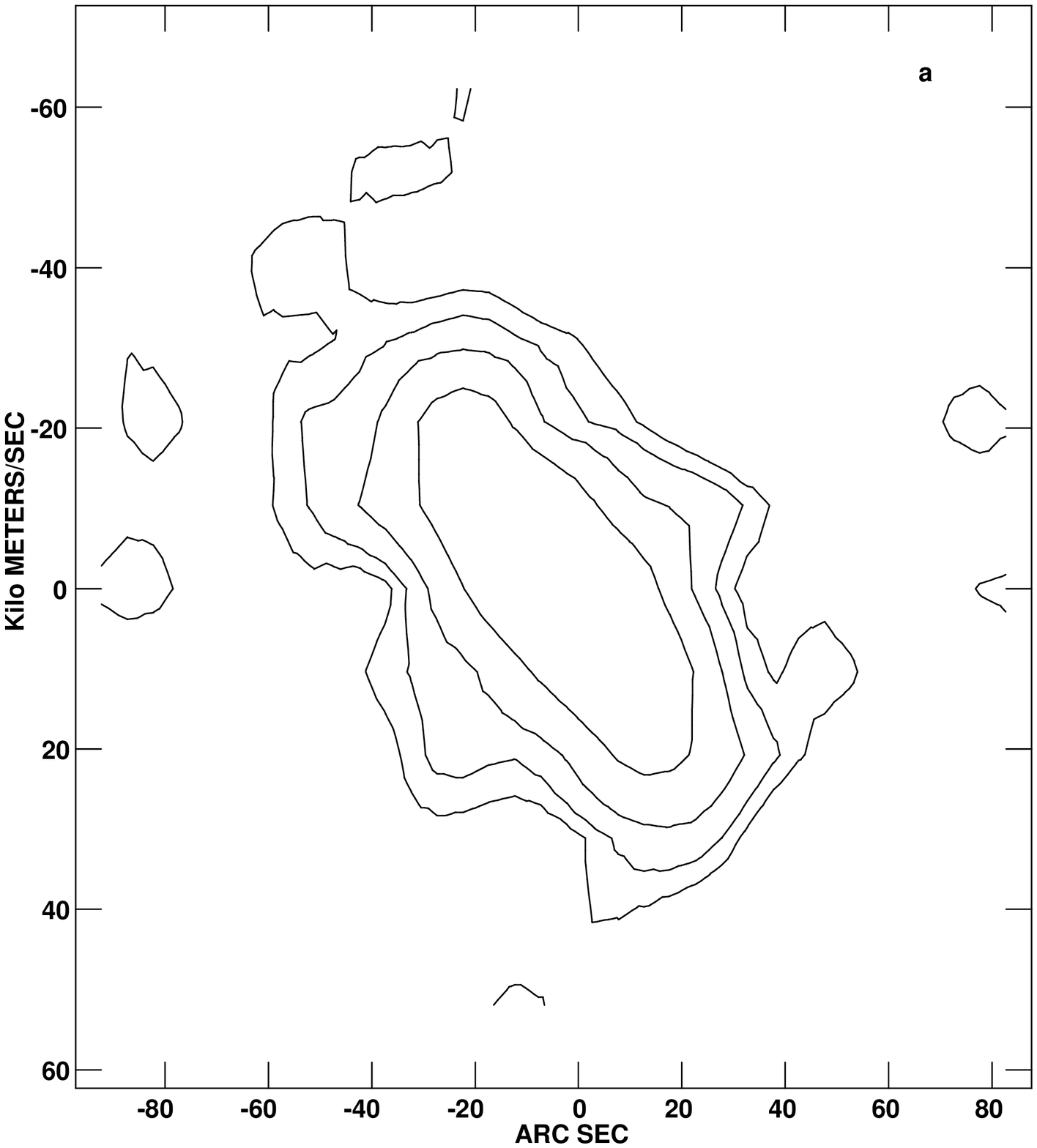}{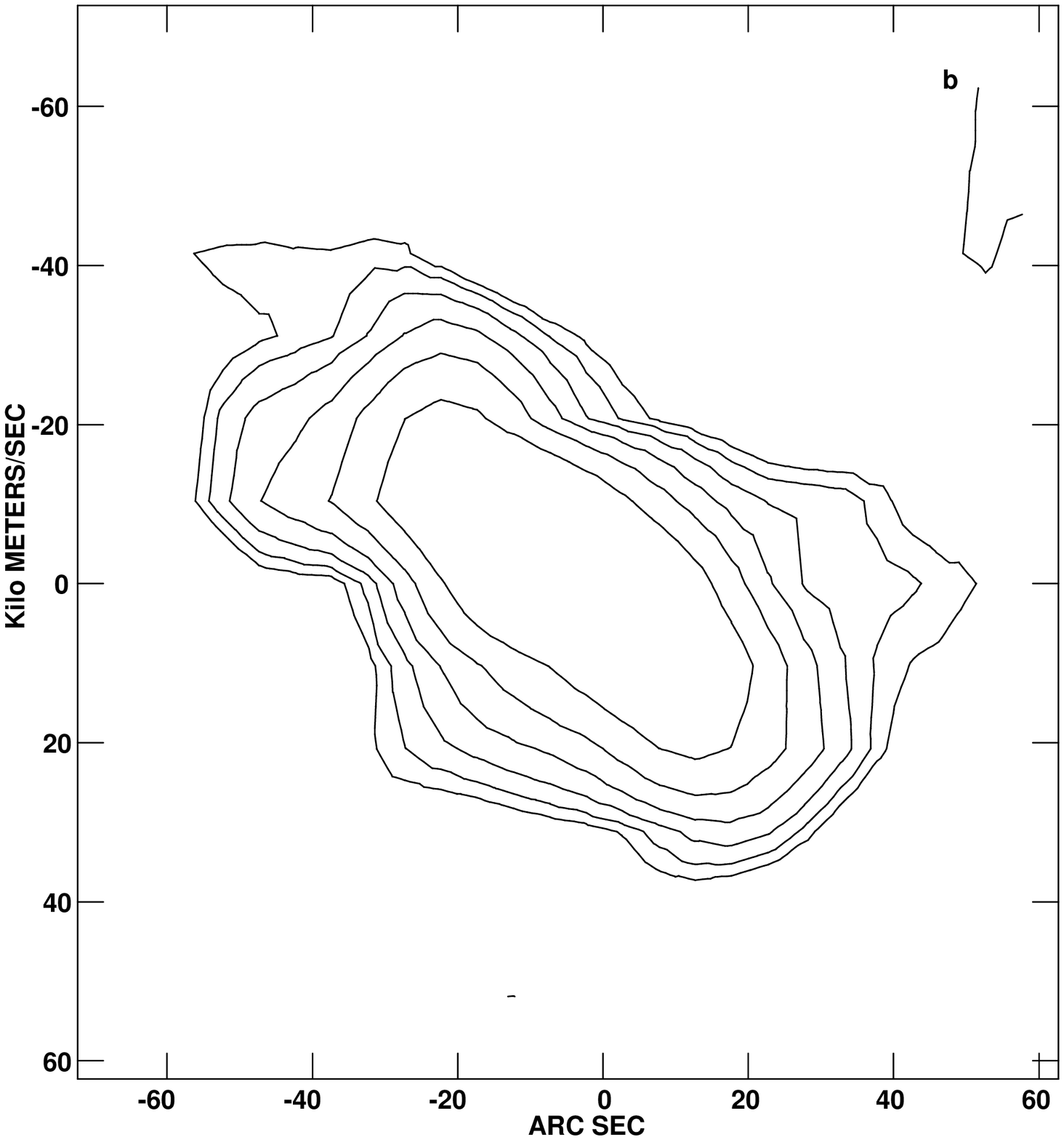}
\caption{
\label{mcgpv}
Position-velocity diagrams for (a) a cut along the major axis of MCG+00-32-16, and (b) for the galaxy summed along the minor axis.
Contours are drawn at 1.0, 2.0, 4.0 and 8.0 mJy ${\rm Bm}^{-1}$ in (a), and at 10.0, 15.0, 22.5, 33.8, 50.6 and 75.9 mJy ${\rm Bm}^{-1}$ in (b).}
\end{figure}

In Fig. \ref{mcgpv} we show position-velocity diagrams for a cut along the major axis of the galaxy and for the galaxy projected onto its major axis (i.e., summed along the minor axis).
These confirm the solid-body nature of the rotation.

\begin{figure}
\plotone{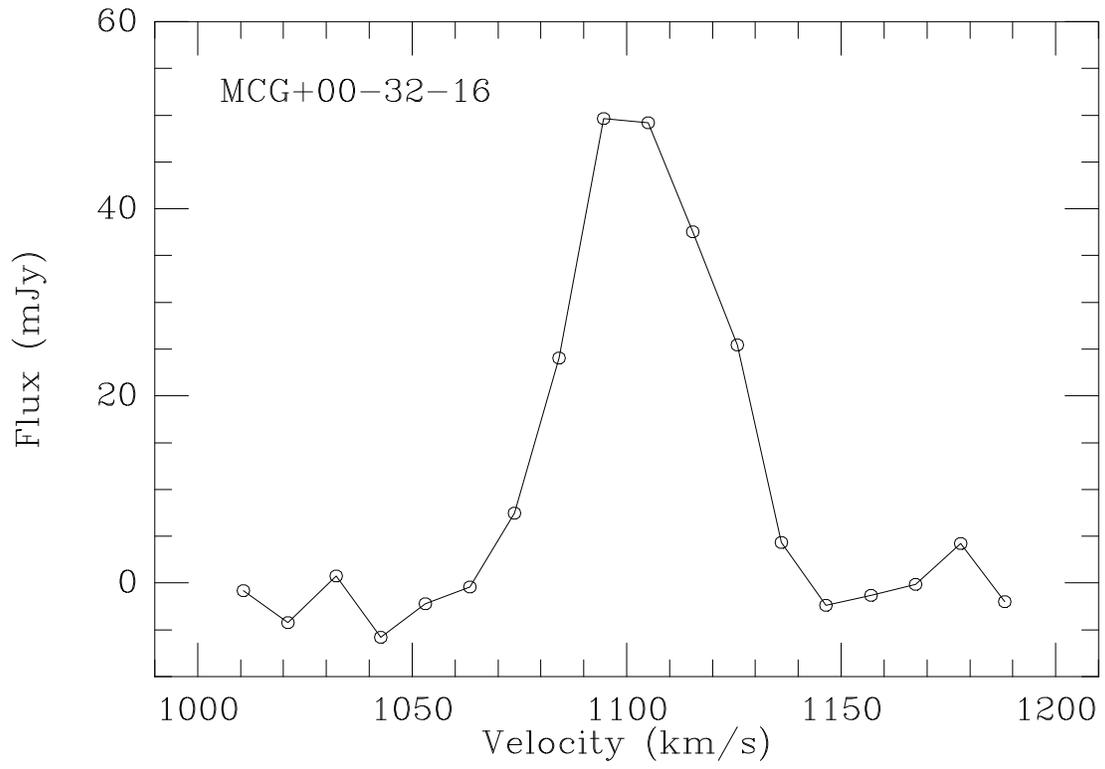}
\caption{
\label{mcgspec}
Integrated spectrum of MCG+00-32-16.}
\end{figure}

The integrated spectrum of the galaxy is shown in Fig. \ref{mcgspec}.
From it, we can estimate a profile width at 50\% of the peak and a systemic velocity at the midpoint between the two sides at the 50\% points:  $\Delta V_{50} = 42$ km ${\rm s}^{-1}$ and $V_{sys} = 1106$ km ${\rm s}^{-1}$.
These agree well within the uncertainties with those measured by Garcia et al. \markcite{GBGGP93}(1993) at Nan\c{c}ay.

\begin{figure}
\epsscale{0.75}
\plotone{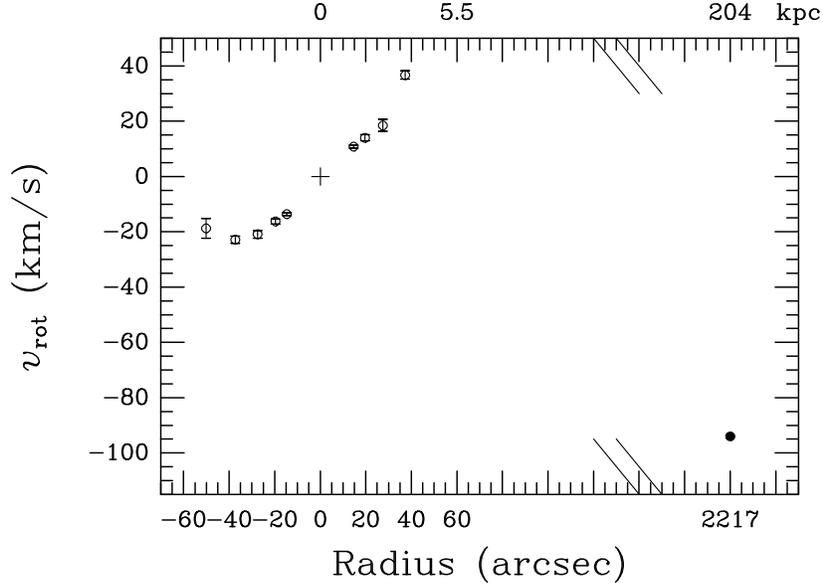}
\caption{
\label{mcgrotcur}
The rotation curve of MCG+00-32-16, with the approaching half of the galaxy shown with negative positions and velocities, positive values shown for the receding half.
Fixed inclination and position angle were used for all rings, which are all centered on the optical nucleus of the galaxy.
The velocity, relative to the galaxy's center, of the lowest redshift Ly$\alpha$ absorber in the spectrum of 3C$\;$273 is shown with a solid point at its position, $2217\arcsec$ along the receding major axis.}
\end{figure}

The AIPS task GAL allows us to determine the rotation velocities within annuli centered on the optical nucleus of the galaxy.
The nearly solid-body rotation curve of the galaxy makes it impossible for us to determine a kinematical inclination; we are forced to rely on the optical axes to determine $i = 40\arcdeg$.
We keep that and the position angle of the \ion{H}{1} distribution, $\approx 132\arcdeg$ (or $312\arcdeg$ for the receding end; this differs slightly from the position angle of the optical isophotes), fixed for all rings.
In Fig. \ref{mcgrotcur} the resulting rotation curve is plotted, with negative positions and velocities corresponding to the approaching (SE) side and positive values to the receding side.
Clearly, no continuation of the rotating disk could account for the absorber since its velocity difference is opposite the sense of rotation of the galaxy.
There is a suggestion of a warp, with the rotation curve turning over on the approaching side but continuing to rise steeply on the receding side, but the statistical significance is low.
If we assume that the last point on each side is unduly influenced by non-circular motions and uncertainty in its inclination, we might take $v_{rot} = 20.5$ km $\rm s^{-1}$ as the maximum rotation velocity and assume that that value holds out to the maximum radius at which significant flux is observed, $r = 50\arcsec$.
Then the dynamical mass of the galaxy is $M_{dyn} = ( v_{rot}^{2} + 3 \sigma_{z}^{2} ) r / G = 8.2 \times 10^8 (d / 19 \,{\rm Mpc})\:\rm M_{\sun}$, where we have included the contribution due to random velocities (Hoffman et al. \markcite{HSFRWH96}1996).

\subsection{CCD surface photometry}

\begin{figure}
\plottwo{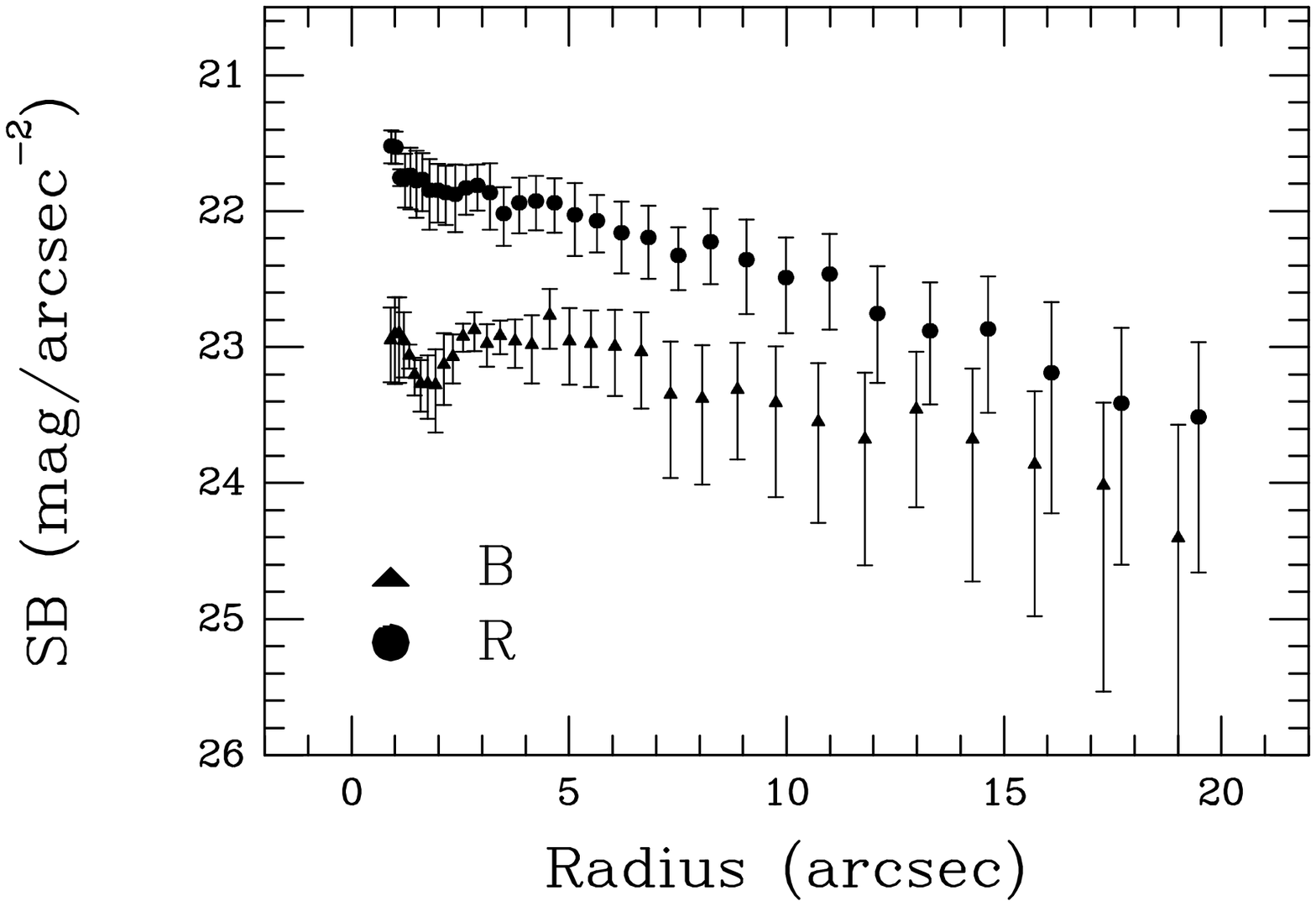}{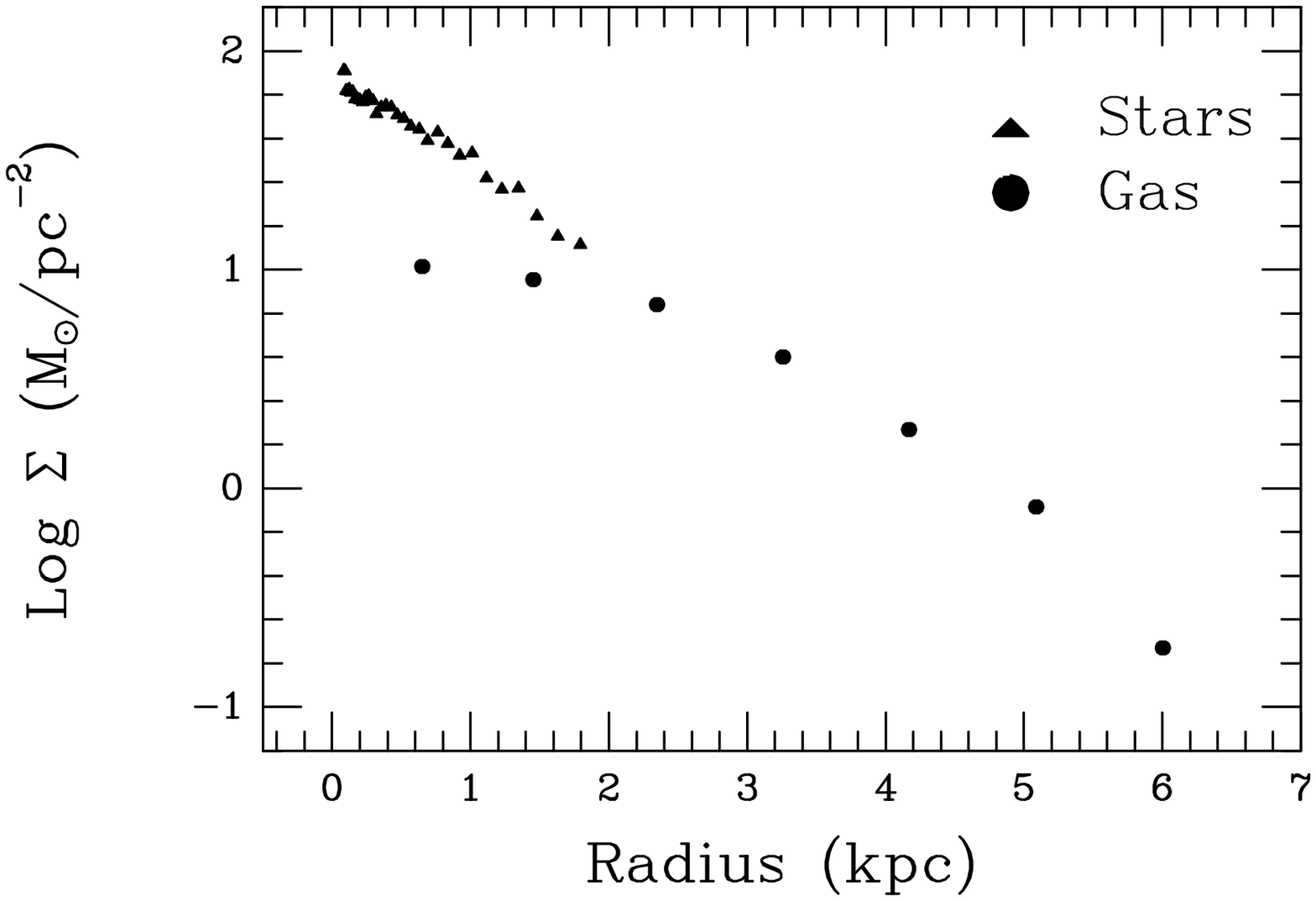}
\caption{
\label{BRprofs}
({\em left panel})  Surface brightness profiles in B and R for MCG+00-32-16 from CCD imaging at Table Mountain Observatory.
The B profile is shown with triangles, R with circles.
({\em right panel})  Mass surface densities in stars (triangles) and gas (circles).
We have assumed $M/L_R = 1\:{\rm M}_{\sun}/{\rm L}_{\sun}$ for the purposes of this plot, and have included neutral hydrogen and primordial helium gas but have neglected molecular and ionized gas.}
\end{figure}

CCD images in B and R were taken under photometric conditions with
the 24-inch telescope at the Table Mountain Observatory of Jet Propulsion 
Laboratory on May 16, 1994 (UT).  
The resolution is limited by a poor seeing of about $3.5\arcsec$ (FWHM).   
The total integration times are 1200 seconds in R and 2400 seconds in B.  
The final magnitudes were tied to the Johnson B and Cousins R systems via observing standard stars in Landolt \markcite{La83}(1983).   
Surface photometry analysis was performed on each of the images following the prescription given by Lu et al. \markcite{LHGRL93}(1993).  
The resulting surface brightness profiles are shown in Fig. \ref{BRprofs}.  
In Table \ref{CCDtab}, we give total magnitudes in B and R, the isophotal
diameters at B$\,$25 and R$\,$24.5$\,$mag$\,$arcsec$^{-2}$, as well as
the following parameters for the galaxy disk:  the mean ellipticity, 
the position angle, the exponential scale length $r_s$ as defined in
$\mu(r) = \mu_0 + 1.086 r/r_s$, and the central surface brightness 
$\mu_0$.   
Since the R-band image is deeper,  these disk parameters are
derived from the fit to the R data alone.    
The quoted statistical 
r.m.s.~uncertainties in Table \ref{CCDtab} are our best estimates.

\subsection{Rotation curve decomposition}

Plots of mass surface densities in stars (assuming $M/L_R = 1\:{\rm M}_{\sun}/{\rm L}_{\sun}$ for the purposes of this figure) and gas (including \ion{H}{1} and primordial helium but neglecting ionized and molecular gas) are given in the right-hand panel of Fig. \ref{BRprofs}.
The stellar disk appears to fall exponentially over the entire range of radii for which surface brightnesses could be measured.
The \ion{H}{1} profile flattens out at small radii, but at large radii it continues the exponential distribution of the stellar disk.
It is reasonable to suppose that the baryonic (stars plus gas) disk falls exponentially from the center to the outermost observed \ion{H}{1} contour.
The scalelength of the stellar disk is $9.3\arcsec$, or 0.86 kpc; that of the \ion{H}{1} is 1.73 kpc if the fit includes all points at radii $\lesssim 5$ kpc, or 1.28 kpc if the fit is restricted to the first 4 points beyond the edge of the stellar disk.

\begin{figure}
\plotone{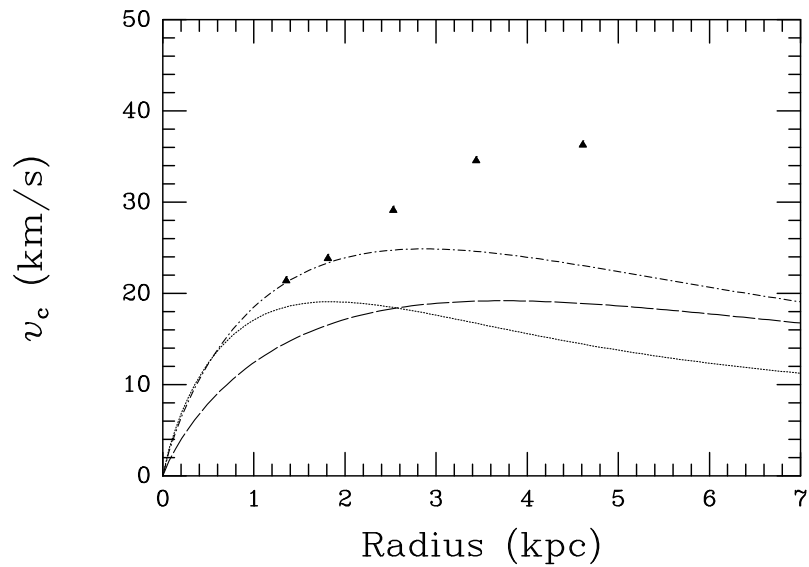}
\caption{
\label{rcfit}
Contributions to the rotation curve of MCG+00-32-16 from luminous matter.
$(M/L)_R = 0.4$ is assumed and is evidently close to the maximum disk mass-to-light ratio.
An exponential fit to the stellar disk is shown with a dotted curve, an exponential fit to the gas ({\sc Hi} plus He I)  with a dashed curve, and an exponential fit to the sum of the two is shown with a dash-dot curve.
The observed ``rotation'' velocity $v_c$, shown with solid triangles, includes a contribution from random motions as discussed in the text.}
\end{figure}

If random motions are not included in the analysis, we cannot fit the observed rotation curve at all unless the galaxy is very close to face-on ($i < 15 \arcdeg$) or the mass-to-light ratio in the R band, $(M/L)_R$, is miniscule.
Therefore we take the dynamically relevant velocity to be $v_c^2 = v_{rot}^2 + 3 \sigma^2$ as discussed in Hoffman et al. \markcite{HSFRWH96}(1996).
A maximum disk hypothesis then gives $(M/L)_R \sim 0.4$; the decomposition is shown in Fig. \ref{rcfit}.
There is the usual need for a dark matter halo beyond $r \gtrsim 2$ kpc.
Incidentally, that same mass-to-light ratio gives the best single exponential fit to the combined stars+gas disk, with a scalelength of $1.33 \pm 0.02$ kpc.

\section{\label{discuss}Discussion}

\begin{figure}
\plotone{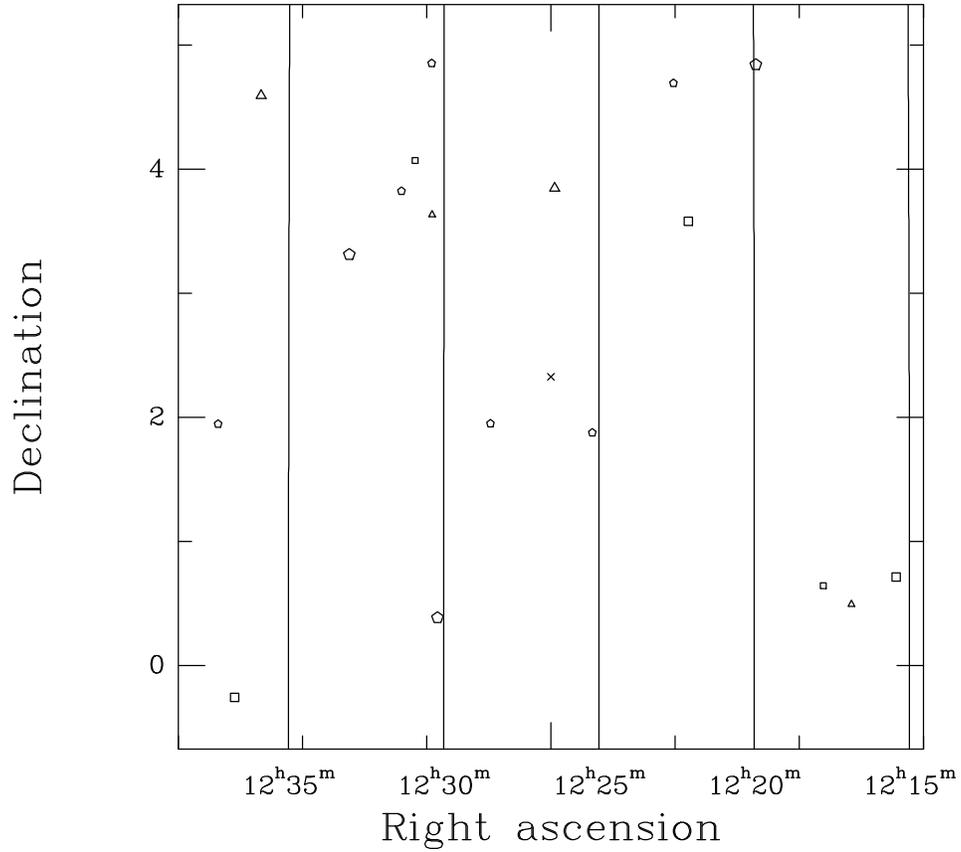}
\caption{
\label{groupmap}
Map of the sky showing all known galaxies in the heliocentric velocity range $700 - 1300\;{\rm km}\;{\rm s}^{-1}$.
The symbol ``$\times$'' marks the line-of-sight to 3C$\;$273.
Triangles denote galaxies with $700 \leq v_{\sun} < 900\;{\rm km\;s}^{-1}$, squares those with $900 < v_{\sun} \leq 1100\;{\rm km\;s}^{-1}$, and pentagons those with $1100 < v_{\sun} \leq 1300\;{\rm km\;s}^{-1}$.
Larger symbols denote galaxies with greater blue luminosity.}
\end{figure}

The galaxy MCG+00-32-16 is about as ``regular'' as an irregular galaxy gets.
With $L_B = 2.02 \times 10^{8}\:{\rm L}_{\sun}$, $L_R = 3.45 \times 10^{8}\:{\rm L}_{\sun}$ (corresponding to a stellar mass $\sim 1.4 \times 10^8\,{\rm M}_{\sun}$), $M_{HI} = 2.0 \times 10^{8}\:{\rm M}_{\sun}$, $M_{dyn} = 12.0 \times 10^{8}\:{\rm M}_{\sun}$, blue optical radius $R_{25} = 2.5$ kpc, and $V_c = \sqrt{v_{rot}^2 + 3 {\sigma}_z^2} = 33.5$ km ${\rm s}^{-1}$ (all assuming a 19 Mpc distance), MCG+00-32-16 is slightly high in $L_B$ as compared to radius, dynamical mass or \ion{H}{1} mass, but well within the scatter of other irregular galaxies that have been mapped in \ion{H}{1} (Salpeter \& Hoffman \markcite{SH96}1996).
There are no signs of out-lying \ion{H}{1} clouds to a column density limit of about $1.1 \times 10^{20}$ atoms ${\rm cm}^{-2}$.
We know, however, that there is low column density gas at a radius $2217\arcsec$, or $204 (d / 19\,{\rm Mpc})$ kpc, more or less along the major axis toward the NW, as indicated by the arrow drawn on Fig. \ref{mcgtothi}.
This is the position of the line-of-sight to 3C$\;$273, where HST observations found a Ly$\alpha$ absorber at heliocentric velocity 1012 km $\rm s^{-1}$ (Bahcall et al. \markcite{BJSHBJ91}1991; Morris et al. \markcite{MWSG91}1991; Morris et al. \markcite{MWD+93}1993). 
Among the galaxies closest in projection to the line-of-sight, MCG+00-32-16 has the closest redshift (1106 km $\rm s^{-1}$) to that of the absorber (Salpeter \& Hoffman \markcite{SH95}1995) while the ``protogalaxy'' \ion{H}{1}$\;$1225+01 (Giovanelli \& Haynes \markcite{GH89}1989), at $2022\arcsec = 186 (d / 19\,{\rm Mpc})$ kpc, is slightly closer in projected distance (redshift 1284 km $\rm s^{-1}$).
The projected distance between MCG+00-32-16 and \ion{H}{1}$\;$1225+01 is $2973\arcsec = 274 (d / 19\,{\rm Mpc})$ kpc.
Fig. \ref{groupmap} shows that the three objects are well separated from all other known galaxies at comparable redshifts in that part of the sky.

Note that, if the absorber were attached to MCG+00-32-16, the absorber must be counter-rotating with respect to the \ion{H}{1} disk (see Fig. \ref{mcgrotcur}).
The velocity difference between either galaxy and the absorber is much greater than the maximum rotation velocity of either galaxy.
Therefore it seems unlikely that the absorber is directly associated with either galaxy.
This should be compared to the findings of Barcons et al. \markcite{BLW95}(1995) for absorbers in the spectra of the quasars 1704$+$6048 and 2135$-$1446; in each of those cases the absorber happened to fall on the side of the major axis where the galaxy rotation creates gas velocity tending toward the absorber's velocity while in our case the absorber falls on the opposite side of the major axis.
Consequently, while the extended disk [type (1)] interpretation was tenable for Barcons et al. (although the galaxy-absorber velocity differences are rather large), it is not tenable for MCG+00-32-16.
With the absorber being closer to one galaxy in velocity and closer to the other in projected distance, all three objects are likely to form one group (possibly with additional unobservable Ly$\alpha$ forest clouds that happen not to fall along the line of sight to the quasar).
There are no other galaxies with $V_{sys}$ in the range 700 to $1300\;$km$\;{\rm s}^{-1}$ within $1.5\arcdeg$ of the 3C$\,$273 sight-line.
Mo et al. (\markcite{MMB94}1994) and Dalcanton et al. (\markcite{DSGSS97}1997a) have suggested that low surface brightness galaxies (of which MCG+00-32-16 and \ion{H}{1}$\;$1225+01 are examples) show slightly less correlation than bright galaxies and are likely to reside preferentially in loose galaxy groups with small velocity dispersion.
The total velocity spread in this group (absorber to \ion{H}{1}$\;$1225+01) is only 272$\;$km$\;{\rm s}^{-1}$, suggesting a smaller velocity dispersion than for typical galaxy groups (Ramella et al. \markcite{RPG97}1997).

Several of the papers cited in the Introduction have discussed low redshift Ly$\alpha$ forest absorbers residing in galaxy groups, either as (a) tidal debris from individual visible galaxies or (b) separate, optically invisible objects, each with its own dark matter halo.
Poor groups with very small total galaxy luminosity are better candidates for (b) than for (a).
The possible fate of gas in a halo of radius $R_{ha}$ with characteristic velocity $V_{circ}$, under the influence of the ubiquitous ionizing UV flux, has been discussed by many authors (at least for isolated halos):
({\em i}) For $V_{circ} \ll 30\;$km$\;{\rm s}^{-1}$, the gas temperature $T \sim 2 \times 10^{4}\;$K imposed by the UV causes the gas to escape from the halo.
In a group environment collisions heat the gas further and it can contribute very little to Ly$\alpha$ absorption (Mulchaey et al. \markcite{MMBD96}1996).
({\em ii}) For $V_{circ} \gg 60\;$km$\;{\rm s}^{-1}$ most, but not all, of the gas collapses to radii much smaller than $R_{ha}$ conserving angular momentum acquired through tidal torquing (see Dalcanton et al. \markcite{DSS97}1997b), forming a disk galaxy with a sharp outer \ion{H}{1} edge (Maloney \markcite{Ma93}1993; Corbelli \& Salpeter \markcite{CS93}1993; Dove \& Shull \markcite{DS94}1994).
The small neutral fraction of gas between this edge and $R_{ha}$ presumably supplies the Ly$\alpha$ forest of type (1).
({\em iii}) For dwarf irregular galaxies with $V_{circ}$ at or just above $60\;$km$\;{\rm s}^{-1}$, there are various possibilities for removing much of the inner gas to the halo leaving very little stellar luminosity (faded dwarfs or Cheshire Cats).
These nearly invisible galaxies are good candidates for the type (2) Ly$\alpha$ forest (Babul \& Rees \markcite{BR92}1992; Salpeter \markcite{Sa93}1993; Vedel et al. \markcite{VHS94}1994; Efstathiou \markcite{Ef95}1995; Salpeter \& Hoffman \markcite{SH95}1995; Quinn et al. \markcite{QKE96}1996; Babul \& Ferguson \markcite{BF96}1996; Shull et al. \markcite{SSP96}1996).

({\em iv}) The transition region $V_{circ} \sim (30\;{\rm to}\;60)\;$km$\;{\rm s}^{-1}$ has become an even more attractive candidate for the type (2) forest as a result of some recent calculations (Thoule \& Weinberg \markcite{TW96}1996; Tegmark et al. \markcite{TSRBAP97}1997; Kepner et al. \markcite{KBS97}1997).
Over an appreciable range of values of $V_{circ}$ most of the gas is able to stay in thermal and Virial equilibrium while filling the whole halo volume.
Shull et al. (\markcite{SSP96}1996) have shown that a conservative extrapolation of the galaxy luminosity function (Marzke et al. \markcite{MHG94}1994) could give enough such ``failed dwarfs'' for the Ly$\alpha$ forest if $R_{ha} \sim 100\;$kpc, contributing a dark matter density parameter $\Omega_{fd} \sim 0.03 ( 100\;{\rm kpc} / {R_{ha}} )$.
If $R_{ha}$ scales with the optical radius for disk galaxies, then $R_{ha} \propto V_{circ}^{1.4}$ (Salpeter \& Hoffman \markcite{SH96}1996) and slightly smaller values of $R_{ha}$ may be more likely.
There is ongoing controversy on the number density of undetected low surface brightness dwarfs (Briggs \markcite{Br97}1997; Dalcanton et al. \markcite{DSGSS97}1997a), but there may be a sufficient number of failed dwarfs even for smaller $R_{ha}$.
Calculations are still needed to see whether the intra-group gas pressure can confine gas in the halo of the failed dwarf for even smaller values of $V_{circ}$.

If Ly$\alpha$ forest absorption in galaxy groups is partly due to tidal debris and partly due to failed dwarf galaxies, one might expect a double-humped distribution of metal abundances (tidal debris having the larger abundance).
Better statistics are needed in general and for the 3C$\;$273 absorber in particular there are only rather high upper limits (Weymann et al. \markcite{WRWMH95}1995; Brandt et al. \markcite{Br+97}1997).

\section{\label{conclusions}Conclusions and summary}

Neutral hydrogen synthesis mapping of the irregular galaxy MCG+00-32-16, closer than any other galaxy in position-redshift space to the lowest redshift absorber along the 3C$\;$273 line-of-sight, reveals that it is rotating in the wrong sense for the absorber to have anything to do with an extended disk of the galaxy, even though the absorber is only 204 kpc distant in projection from the galaxy and has velocity difference only 94 km ${\rm s}^{-1}$.
Neither can an extended ``disk'' around the next closest (by a small margin) galaxy, \ion{H}{1}$\;$1225+01 (Giovanelli et al. \markcite{GWH91}1991), account for the absorber.
This system therefore runs counter to the interpretation [our Type (1)] presented by Barcons et al. \markcite{BLW95}(1995) for two otherwise similar cases in which the rotation turned out to be in the other sense.

Measurement of the luminosity profile in B and R, along with the \ion{H}{1} column density profile, allow us to assess the contributions of the gas and stars to the observed rotation curve.
The ``maximum disk'' fit requires only a modest $(M/L)_R \sim 0.3$ for the stellar disk, but dark matter is required outside 1.5 scalelengths as in most irregular and spiral galaxies.
Our measurements of total luminosity give $L_B = 2.02 \times 10^{8}\:{\rm L}_{\sun}$, $L_R = 3.45 \times 10^{8}\:{\rm L}_{\sun}$, and $M_{HI} = 2.0 \times 10^{8}\:{\rm M}_{\sun}$.
The 50\% profile width $\Delta V_{50} = 42\;{\rm km\;s}^{-1}$ and blue optical radius $R_{25} = 2.5$ kpc lead to $M_{dyn} = 12.0 \times 10^{8}\:{\rm M}_{\sun}$ (all assuming a 19 Mpc distance).

The failure of the extended disk hypothesis makes it more likely that MCG+00-32-16, \ion{H}{1}$\,$1225+01 and the 3C$\,$273 absorber form part of a sparse galaxy group.
The lack of evidence for tidal debris in the 3C$\;$273 system has led us to consider other models for Ly$\alpha$ gas in a galaxy group.
In particular, the ``failed dwarf'' model (see Shull et al. \markcite{SSP96}1996), assisted by intra-group gas pressure, may be attractive for absorbers in loose galaxy groups.
Heavy element abundances would help to discriminate between fds and tidal debris, but the column densities of the 3C$\;$273 low redshift absorbers are too low to permit the necessary measurements with existing (or envisioned) instruments.

\acknowledgments

We thank J. Young of JPL Table Mountain Observatory for help with the CCD observations,
and S. Morris, J. Webb, and R. Weymann for illuminating discussions.
Perceptive comments by the anonymous referee helped us to improve the presentation of the data and reminded us of some prior discussions of fds.
This work was supported in part by US National Science Foundation grants AST-9015181 and AST-9316213 at Lafayette College; in part by Jet Propulsion Laboratory, California Institute of Technology, under a contract with the National Aeronautics and Space Administration; and in part by the J.G. White Memorial Account at Cornell University.
The Digitized Sky Surveys were produced at the Space Telescope Science Institute under U.S. Government grant NAG W-2166.

\begin{table}
\dummytable\label{obsdet}
\end{table}

\begin{figure}
\plotone{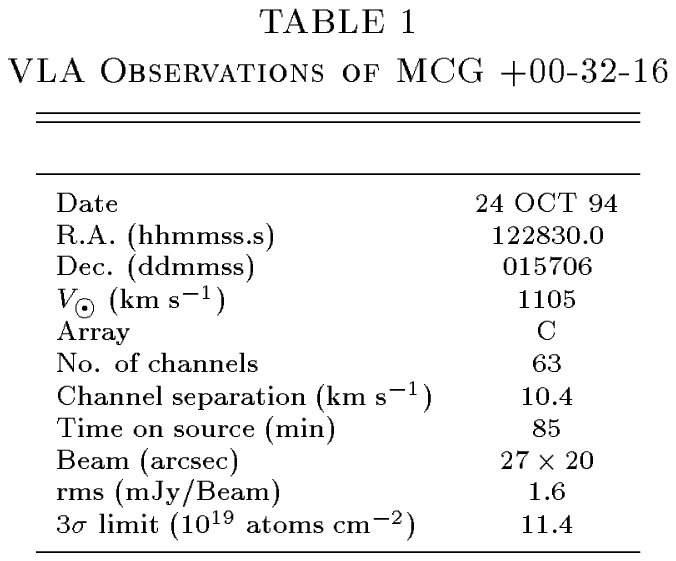}
\end{figure}

\begin{table}
\dummytable\label{CCDtab}
\end{table}

\begin{figure}
\plotone{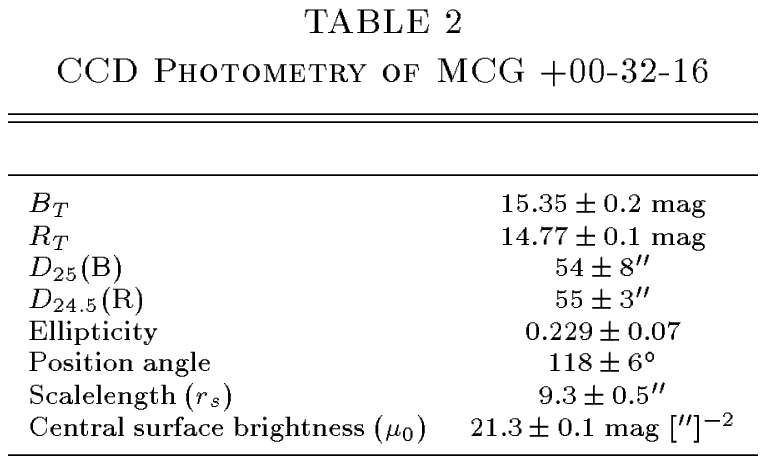}
\end{figure}

\end{document}